\documentclass[a4paper]{jpconf}

\usepackage{amsmath}
\usepackage{bm}
\usepackage{graphicx}
\usepackage{epstopdf}

\begin{document}
\title{Axion cold dark matter revisited}

\author{L Visinelli$^{\dag}$\footnote{Speaker. Talk based on L Visinelli, P Gondolo, Phys.\ Rev.\ D {\bf 80}, 035024 (2009).}, P Gondolo$^{\ddag}$}

\address{Department of Physics, University of Utah, 115 S 1400 E
rm.201, Salt Lake City, UT 84102, USA}

\ead{$^{\dag}$ visinelli@utah.edu}
\ead{$^{\ddag}$ paolo@physics.utah.edu}

\begin{abstract}
We study for what specific values of the theoretical parameters the axion can form the totality of cold dark matter. We examine the allowed axion parameter region in the light of recent data collected by the WMAP5 mission plus baryon acoustic oscillations and supernovae \cite{komatsu}, and assume an inflationary scenario and standard cosmology. We also upgrade the treatment of anharmonicities in the axion potential, which we find important in certain cases. If the Peccei-Quinn symmetry is restored after inflation, we recover the usual relation between axion mass and density, so that an axion mass $m_a =(85\pm3){\rm~\mu eV}$ makes the axion 100$\%$ of the cold dark matter. If the Peccei-Quinn symmetry is broken during inflation, the axion can instead be 100$\%$ of the cold dark matter for $m_a < 15{\rm~meV}$ provided  a specific value of the initial misalignment angle $\theta_i$ is chosen in correspondence to a given value of its mass $m_a$. Large values of the Peccei-Quinn symmetry breaking scale correspond to small, perhaps uncomfortably small, values of the initial misalignment angle $\theta_i$.
\end{abstract}

\section{Introduction}

About 84$\%$ of the non-relativistic matter in the Universe is in the form of cold dark matter (CDM)\cite{komatsu}, whose nature is still to be discovered. One of the most promising hypothetical candidates that could account for the CDM observed is the axion \cite{weinberg, peccei}.

The properties of the axion as the CDM particle have been studied in various papers \cite{preskill, beltran}. We examine the possibility that the axion accounts for 100$\%$ CDM in the light of the WMAP5 mission, baryon acoustic oscillations (BAO) and supernovae (SN) data. We also upgrade the treatment of anharmonicities in the axion potential, which we find important in a specific region of the parameter space. The axion parameter space is described by three quantities, the PQ energy scale $f_a$, the Hubble parameter at the end of inflation $H_I$ and the axion initial misalignment angle $\theta_i$.

\section{Properties of the axion}

The solution to the strong-CP problem by Peccei and Quinn \cite{peccei} introduces a new $U(1)_{PQ}$ symmetry in the theory of strong interactions that spontaneously breaks at a temperature of order the PQ energy scale, $T \sim f_a$. The axion is the pseudo Nambu-Goldstone boson originating from the breaking of $U(1)_{PQ}$. We denote by $a(x)$ the axion field and $\theta(x) = a(x)/f_a$ the misalignment field. At a temperature of order the QCD transition $T \sim \Lambda_{QCD}$ an effective potential $V(\theta)$ arises through instanton effects \cite{gross} given by
\begin{equation}
V(\theta) = m_a^2(T)\,f_a^2\,\left(1-\cos\theta\right).
\end{equation}
The temperature-dependent axion mass $m_a(T)$ is given by
\begin{equation} \label{axion_mass}
m_a(T) = \begin{cases}
m_a \,b\, \left(\frac{\Lambda}{T}\right)^4 & T > \Lambda_{QCD},\\
m_a & T < \Lambda_{QCD}.
\end{cases}
\end{equation}
We set $\Lambda_{QCD} = 200$MeV, $b=0.018$ \cite{beltran}. Setting $f_{a,12} = f_a/10^{12}{\rm GeV}$, the axion mass at zero temperature is
\begin{equation} \label{axion_mass_zero}
m_a = 6.2 {\rm \mu eV}\,f_{a,12}^{-1}/N.
\end{equation}
The integer $N$ represents the $U(1)_{PQ}$ color anomaly index; here $N=1$.

\section{Axion and cosmology}

Setting $T_{GH} = H_I/2\pi$, axion cosmology is different if the PQ symmetry breaks after inflation, $f_a < T_{GH}$ (Scenario I), or during inflation, $f_a > T_{GH}$ (Scenario II). Over a Hubble volume, $\theta_i$ takes different values in Scenario I, while in Scenario II $\theta_i$ takes only one random value.

The subsequent evolution of the zero mode of the dynamical field $\theta(x)$ in a flat Friedmann-Robertson-Walker metric is given by
\begin{equation}\label{eq_motion}
\ddot{\theta} + 3H(T)\,\dot{\theta}
+\frac{1}{f_a^2}\frac{\partial V(\theta)}{\partial\, \theta} = 0.
\end{equation}
Here a dot indicates a derivative with respect to time and $H(T)$ is
the Hubble expansion rate. For $T \gg \Lambda_{QCD}$ the misalignment field is frozen at a constant value, $\theta \equiv \theta_i = {\rm const}$, where $\theta_i$ is the initial misalignment angle. For $m_a(T) \approx 3H(T)$ the axion field begins to oscillate in the $V(\theta)$ potential. The axion energy density-to-critical density is computed as
\begin{equation} \label{standarddensity}
\Omega^{mis}_a h^2 =
\begin{cases}
0.236\langle\theta_i^2 f(\theta_i)\rangle\, f_{a,12}^{7/6}, & f_a < \hat{f}_a,\\
0.0051\langle\theta_i^2 f(\theta_i)\rangle\, f_{a,12}^{3/2}, & f_a > \hat{f}_a,\\
\end{cases}
\end{equation}
with $\hat{f}_a =  9.91\times 10^{16}{\rm GeV}$. The function $f(\theta_i)$ accounts for anharmonicities in the axion potential, i.e., deviations from a pure harmonic potential when $\theta_i$ is no longer smaller than one. We look for an analytic function $f(\theta_i)$ that interpolates between previous numerical results \cite{turner}. We take
\begin{equation} \label{f_theta}
f(\theta_i) = \left[\ln\left(\frac{e}{1-\theta_i^2/\pi^2}\right)\right]^{7/6}.
\end{equation}
The function $f(\theta_i)$ in eq.(\ref{f_theta}) shows the correct asymptotic behaviors for $\theta_i\to\pi$ and $\theta_i\to0$.

\section{Results}

Figure 1 shows the region of the axion parameter space in which the axion is 100$\%$ CDM, or $\Omega_a = \Omega_{CDM}$, where $\Omega_{CDM}$ is the CDM density in units of the critical density. The PQ scale $f_a$ is bounded from below by measurements on white dwarfs cooling time, $f_a > 4\times 10^8$GeV, while the Hubble rate at the end of inflation $H_I$ is bounded from above by non-detection of tensor modes, $H_I < 6.3\times 10^{14}$GeV. For $f_a < T_{GH}$ (Scenario I), the axion can be 100$\%$ CDM only if $f_a = (7.27 \pm 0.25)\times 10^{10}{\rm ~GeV}$. The function $f(\theta_i)$ in eq.(\ref{f_theta}) affects the results by increasing the value of $\langle\theta_i^2 f(\theta_i)\rangle$. In Scenario II, $f_a > T_{GH}$, the parameter space on the top left is bounded by non-detection on axion isocurvature fluctuations that yield $f_a < 4.17 \times 10^{-5}\,H_I\,\theta_i$. The upper kink in the leftmost boundary is due to the change in the dependence of the axion mass with temperature, eq.(\ref{axion_mass}), while the lower kink is due to the presence of anharmonicities, eq.~(\ref{f_theta}). In the allowed region of Scenario II, for a given $f_a$ the axion is 100$\%$ CDM only if a correspondent value $\theta_i = \theta_i(f_a)$ is selected. This value of $\theta_i$ does not depend on $H_I$. We show the function $\theta_i  = \theta_i(f_a)$ in Figure 2, where we see that a value $\theta_i < \theta_i(f_a)$ for a given $f_a$ makes the axion population sub-dominant compared to the CDM observed, while a value $\theta_i > \theta_i(f_a)$ is over abundant. We included the region excluded by ADMX1 axion direct search and future reaches by the Planck satellite and future cavity searches.

\begin{figure}[h!]
\begin{minipage}[b]{0.5\linewidth}
\centering
\includegraphics[width=8.5cm]{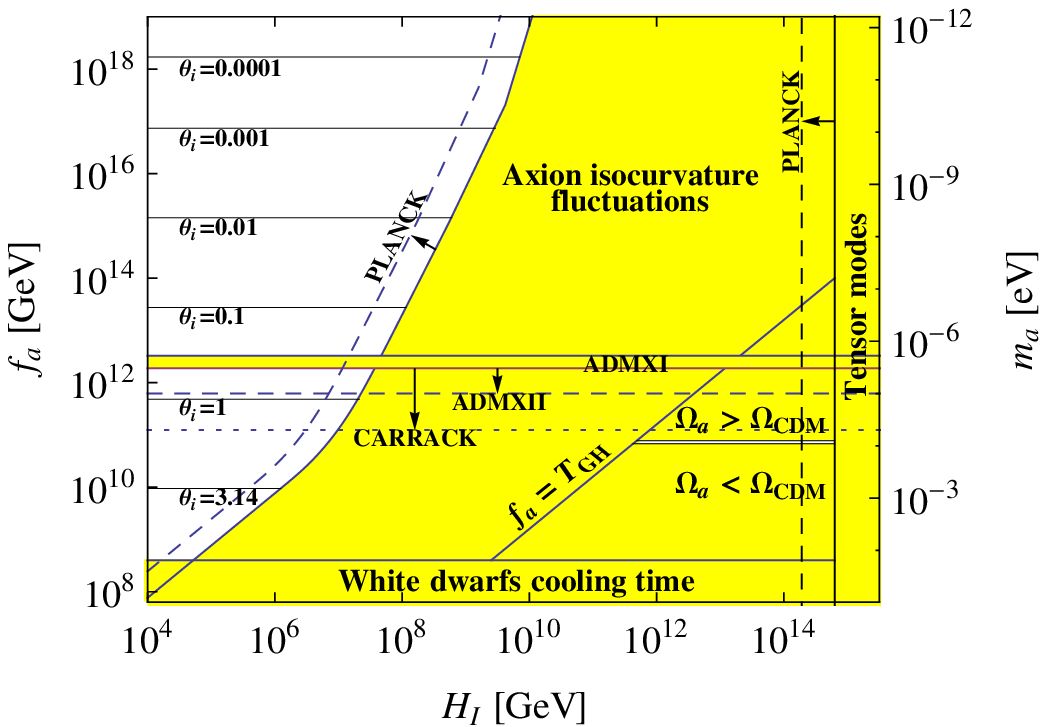}
\caption{\footnotesize Region of axion parameter space where the axion is 100\% of the cold dark matter. The axion mass scale on the right corresponds to eq.~(\ref{axion_mass_zero}) with $N=1$. When the PQ symmetry breaks after inflation ($f_a < H_I/2\pi$), the axion is 100$\%$CDM if $f_a =  (7.27 \pm 0.25)\times 10^{10}{\rm ~GeV}$, or $m_a=(85\pm3){\rm ~\mu eV}$, which is the narrow horizontal window shown on the right. If the axion is present during inflation ($f_a > H_I/2\pi$), axion isocurvature perturbations constrain the parameter space to the region on the top left, which is marked by the values of $\theta_i$ necessary to obtain 100\% of the CDM density. Other bounds indicated in the figure come from observations of white dwarfs cooling times and the non-observation of tensor modes in the CMB spectrum. Dashed lines and arrows indicate the future reach of the PLANCK satellite and the ADMX and CARRACK cavity searches.}
\end{minipage}
\hspace{0.5cm}
\begin{minipage}[b]{0.5\linewidth}
\centering
\includegraphics[width=7cm,height=6.5cm]{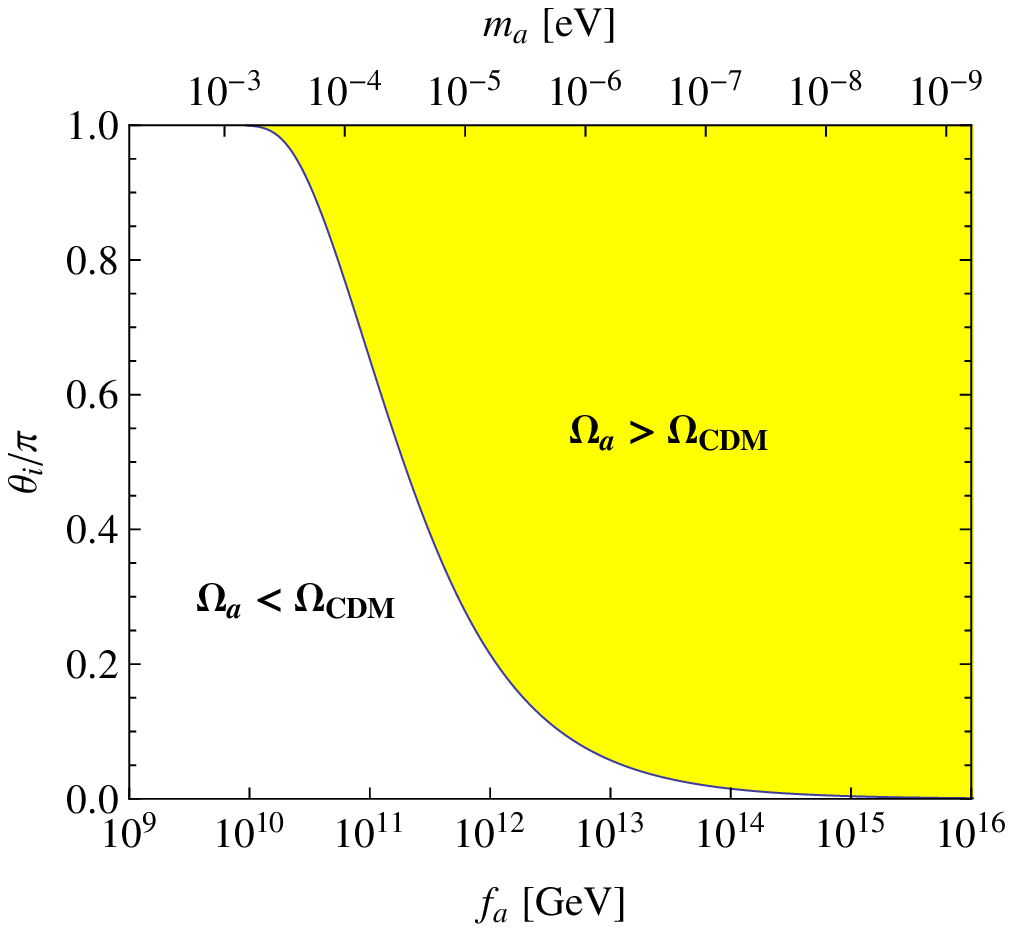}
\caption{\footnotesize The misalignment angle $\theta_i$ necessary for the axion to be 100\% of the cold dark matter in Scenario II ($f_a>H_I/2\pi$), as a function of $f_a$. Above the curve, $\Omega_a > \Omega_{CDM}$. For $f_a > 10^{17}$GeV, $\theta_i \approx 0.001 (f_a/10^{17}\,{\rm GeV})^{-3/4}$.\\
\\
\\
\\
\\
\\
\\
\\
\\
\\
\\
\\}

\end{minipage}
\end{figure}

\end{document}